\documentclass[proceedings]{rmaa}

\def\etal{{\it et al.~}}

\def\eg{{\it e.g.,~}}

\title{Cosmic Rays and Their Radiative Processes in Numerical Cosmology}

\author{Dongsu Ryu
 \affil{Chungnam National University, Korea}
        Francesco Miniati, Tom W. Jones 
  \affil{University of Minnesota, USA}
  \and
        Hyesung Kang
   \affil{Pusan National University, Korea} 
   }

\fulladdresses{
\item Dongsu Ryu: Department of Astronomy and Space Science,
      Chungnam National University, Daejeon, 305-764, Korea
      (ryu@canopus.chungnam.ac.kr)
\item Francesco Miniati, Tom W. Jones: Department of Astronomy,
      University of Minnesota, 116 Church St SE,
      Minneapolis, MN 55455, USA (min,twj@msi.umn.edu)
\item Hyesung Kang: Department of Earth Science,
      Pusan National University, Pusan, 609-735, Korea
      (kang@uju.es.pusan.ac.kr)}

\shortauthor{Ryu et al.}
\shorttitle{Cosmic Rays in Numerical Cosmology}


\abstract{
A cosmological hydrodynamic code is described, which includes a routine
to compute cosmic ray acceleration and transport in a simplified way. The
routine was designed to follow explicitly diffusive acceleration at shocks,
and second-order Fermi acceleration and adiabatic loss in smooth flows.
Synchrotron cooling of the electron population can also be followed. The
updated code is intended to be used to study the properties of nonthermal
synchrotron emission and inverse Compton scattering from electron cosmic
rays in clusters of galaxies, in addition to the properties of thermal
bremsstrahlung emission from hot gas. The results of a test simulation
using a grid of $128^3$ cells are presented, where cosmic rays and magnetic
field have been treated passively and synchrotron cooling of cosmic ray
electrons has not been included.}

\begin{document}

\maketitle

\section{Introduction}

There is growing evidence both in observational and
theoretical studies that cosmic rays may be an important dynamical component, 
which affect the formation and equilibrium of clusters of
galaxies (GCs) and the large scale structure of the universe
(\eg En{\ss}lin \etal 1997). Relativistic cosmic-ray (CR) electrons have
been observed in GCs through their synchrotron emission (\eg Kim \etal 1989).
In addition, many clusters
possess an excess of radiation compared to that
expected from the hot, thermal X-ray emitting Intra Cluster Medium (ICM)
both  in the
extreme ultra-violet (EUV) (\eg Fabian 1996) and in the hard 
X-ray band above $\sim 10$ KeV (\eg Fusco-Femiano \etal 1999). One of 
the mechanisms proposed for the origin of this component is 
inverse-Compton (IC) scattering of cosmic microwave background photons
by CR electrons. 
Based on this interpretation and assuming
diffusive shock acceleration for the origin of CR electrons,
Lieu \etal (1999) concluded that a population of CR proton in equipartition 
of energy with the thermal gas should be present in the Coma cluster. 
However, CR protons have
not been directly observed yet in GCs (\eg Sreekumar \etal 1996).

\section{Code}

As an effort to study the observational signatures and dynamical effects
of CRs in numerical cosmology, we have developed a code which follows
the acceleration and further evolution of CRs along with matter in the
cosmological context. Here, dark matter is treated with the particle-mesh
(PM) method and gas and magnetic field is treated with a second-order
accurate, conservative scheme called the total variation diminishing
(TVD) scheme (Ryu \etal 1993). Special care was taken so that the
code can capture accurately shocks even with very large Mach numbers,
$M \gtrsim 100$. For energetic particle transport we use the
conventional convection-diffusion equation for the momentum
distribution function, $f$, (e.g., Skilling 1975) which follows
spatial and momentum diffusion as well as spatial and momentum advection
of the particles. However, high computational costs prohibit solving this
equation through standard finite difference methods in complex flows. To
circumvent this we use a conservative finite volume approach in the
momentum coordinate, taking advantage of the broad spectral character
expected for $f(p)$. Particle fluxes across momentum bin boundaries are
estimated by representing $f(p)$ as $f(p) \propto p^{-q(p)}$, where $q(p)$
varies in a regular way. Numerically we use the integrated number of
electrons within each bin and the slope, $q$, within each bin. Thus, we
can follow electron spectral evolution in smooth flows with a modest number
of momentum bins. Typically we have used 8 bins to cover energies up to a
few hundred GeV for electrons. In addition, diffusive acceleration of
electrons to GeV energies at shocks is effectively instantaneous within
a dynamical time step. Hence, we assume the analytic, steady, test
particle form for the CR distribution just behind shocks. That is, the
spectrum is a power law with an index, $q = {3 r}/{r - 1}$, where
$r$ is the shock compression ratio. Details for the treatment of CRs
can be found in Jones \etal (1999).

\section{Results}

As a test, a standard cold dark matter (CDM) model universe with
total $\Omega_M=1$ of matter has been simulated in a periodic box with
$(32 h^{-1}{\rm Mpc})^3$ volume using $128^3$ cells and $64^3$ particles
from $z_i=20$ to $z_f=0$. The values of other parameters used are
$\Omega_b=0.06$, $h=1/2$, and $\sigma_8=1.05$. Synchrotron cooling
of CR electrons has NOT been included. Since it is important in real
situations, later simulations with synchrotron cooling are expected
to produce results somewhat different.

With the simulated CRs along with the magnetic field distribution
we can study the properties of nonthermal synchrotron emission and
IC emissivity in CGs, in addition to the properties of thermal
bremsstrahlung emission from hot gas. Figure 1 shows sliced
maps of synchrotron emission (left panel) and bremsstrahlung emission
(right panel) around a cluster identified in the simulation. The
distribution of synchrotron emission roughly follows that of
bremsstrahlung emission, but it shows more fine strutures. This is because
the regions of strong synchrotron emission typically correspond to the
regions of strong magnetic field. Figure 2 plots the average magnetic
field strength, the ratio of CR-proton
to gas pressure, the IC and synchrotron emissivity
from clusters as a function of cluster core temperature. Both
synchrotron and IC emissivity are larger in hotter cluster.
In addition, larger and hotter clusters have stronger magnetic field.
As a result, the slope of synchrotron emissivity is steeper than that
of the IC. With an injection rate $\epsilon_{inj}\sim10^{-2}$ or so,
the top-right panel shows that the CR pressure becomes comparable to
the gas pressure. Hence, we expect that CRs play important dynamical
roles in the formation of large scale structures and clusters.

\begin{figure}
  \begin{center}
    \leavevmode
    \includegraphics[width=7.in]{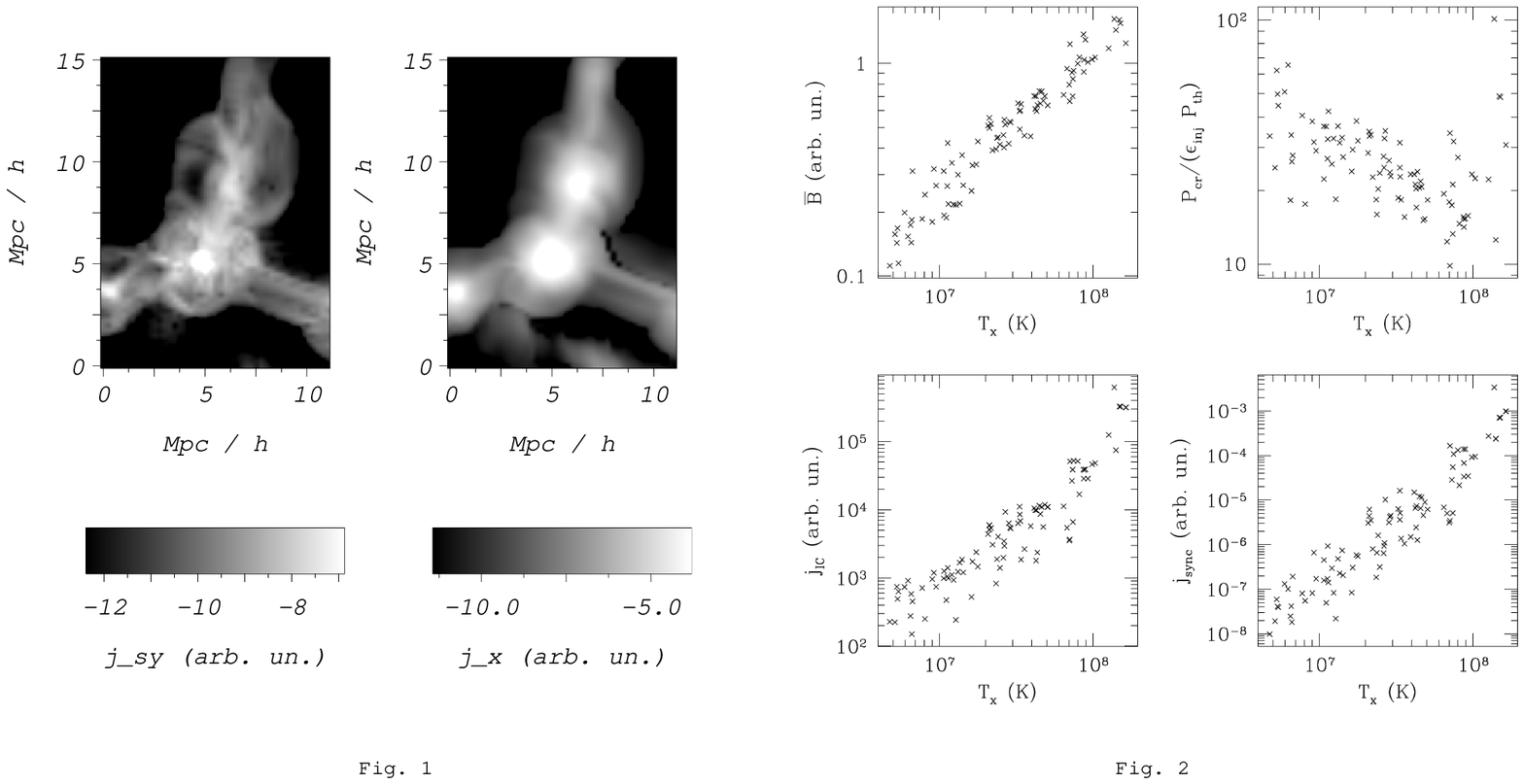}
  \end{center}
\end{figure}

\acknowledgements
DR and HK were supported in part by grant 1999-2-113-001-5 from the
interdisciplinary Research Program of the KOSEF.
FM and TWJ were supported in part by NSF grants AST9616964 and INT9511654, 
NASA grant NAGS-5055 and by the Minnesota Supercomputing Institute.

\end{document}